\documentclass[10pt]{article}
\usepackage{amsmath,amsopn,amscd,amsthm,amssymb,xypic,rotating}
\usepackage{multirow}
\usepackage{libertine}

\def\em{\sl}
\def\emph{\textsl}
\newcommand{\dqt}[1]        {"{#1}"}

\def\textbf{\pmb}

\newcommand{\single}       {\baselineskip 14pt}

\xyoption{all}

\newcommand{\HInsert}[2]   {\immediate\pdfximage width #2  {#1}\pdfrefximage\pdflastximage}

\newcommand{\cb}          {\begin{tabbing}MMMMM\=MM\=MM\=MM\=MM\=MM\=MM\=MM\=MM\=MM\= \kill}
\newcommand{\ce}          {\end{tabbing}}

\title{Structural Characterization of Musical Harmonies}
\author{Mar{\'\i}a Rojo Gonz\'alez \and Simone Santini}
\date{Escuela Polit\'ecnica Superior\\Universidad Aut\'onoma de Madrid}

\begin{document}

\maketitle

\begin{abstract}
  Understanding the structural characteristics of harmony is essential
  for an effective use of music as a communication medium. Of the
  three expressive axes of music (melody, rhythm, harmony), harmony is
  the foundation on which the emotional content is built, and its
  understanding is important in areas such as multimedia and affective
  computing.

  The common tool for studying this kind of structure in computing
  science is the formal grammar but, in the case of music, grammars
  run into problems due to the ambiguous nature of some of the
  concepts defined in music theory.

  In this paper, we consider one of such constructs: modulation, that
  is, the change of key in the middle of a musical piece, an important
  tool used by many authors to enhance the capacity of music to
  express emotions. We develop a hybrid method in which an
  evidence-gathering numerical method detects modulation and then,
  based on the detected tonalities, a non-ambiguous grammar can be
  used for analyzing the structure of each tonal
  component. Experiments with music from the XVII and XVIII centuries
  show that we can detect the precise point of modulation with an
  error of at most two chords in almost 97\% of the cases. Finally,
  we show examples of complete modulation and structural analysis of
  musical harmonies.
\end{abstract}

\single

\section{Introduction}
Music is one of the fundamental media through which the human
experience is created and shared. All cultures, from before the
emergence of Homo Sapiens, have felt the need to emit harmonious
sounds to communicate, to ritualize, to reinforce human bonds. We, as
a species, came to life in a world that was already full of
music. From the traditional dances of the early human culture to the
mating rituals of contemporary young people in clubs and raves, music
has been a constant presence in the life of our species.

The very dawn of western science coincides with the pytha\-gorean
discovery of the relation between music and mathematics: sounds with a
pitch ratio of 2:1 are perceived as the same note in different
octaves, while the ratio 3:2 (the perfect fifth) is pleasing to the
ear and generates a sense of expectation: the first confirmed relation
between mathematics and emotions; affective computing \emph{ante
  litteram}. In the ancient world, the relation between mathematics
and music was so well established that in medieval universities
mathematics consisted of two major branches: geometry and music.

Tonal music, created between the XVI and the XVII century mainly in
Germany, is one of the pinnacles of European civilization, a great
audible mathematical cathedral whose principles are exposed in Bach's
\dqt{The art of the Fugue}. For this is the unique position of music:
it is a very direct emotional medium, one that speaks to us without
the intervening mediation of language (Isaac Asimov once said that
Beethoven is greater that Shakespeare because his work did not need
translation), a truly emotional medium but also, at the same time, one
of the media whose basis is more firmly mathematical.

The idea of using computers to formalize music pre-dates the existence
of actual computers, and can be traced back to an observation by Ada
Lovelace \cite{hooper2012ada, menabrea1842sketch}. For a contemporary
computer scientist, such an attempt is interesting from two points of
view. One the one hand, music is the perfect testbed for analyzing the
formalization of emotionally \dqt{charged} media
\cite{kuo2005emotion}, and the limits of such formalization: music is
mathematical but, like any truly open system, musical creativity is
(fortunately) impossible to formalize. On the other hand, music (and
automatic musical generation) is a fundamental part of any multimedia
experience (try to remove the music from the shower scene in
Hitchcock's \emph{Psycho}) and, in order to understand how to use this
medium properly, one has to realize that music is not merely a
sequence of sounds, but has an harmonic structure. The
characterization of this harmonic structure is the objective of this
paper.

The interest of computing scientists for music goes back to the days
of yore of computing. Surveys on the early work in the field can be
found, for example, in \cite{ames1987automated,papadopoulos1999ai}. 

Work on computer music has developed in various directions. Some work
has concentrated on the sequential aspects of music, the most common
techniques in this direction being Markov models \cite{ames1989markov}
and machine learning \cite{gillick2010machine}. This work has obtained
good results, especially in music composition
\cite{anders2008constraint, bresson2010openmusic,
  keller2007grammatical}, but has fallen short of capturing the
structural aspects of harmony and, in general, of creating a bridge
between mathematical formalization and musical theory as developed,
for example, in \cite{piston1941harmony}. Similar techniques have also
enjoyed considerable success in the area of music retrieval and
recommendation \cite{chai2003music, lew2006content}.

A second line of work has concentrated on the structural
characteristics of harmony. A lot of this work has used formal
grammars as a modeling tool \cite{rohrmeier2007generative,
  salas2010music}. The problem with this approach is that many musical
structures are, as we shall see, strongly contextual, when not
ambiguously defined. The result is that grammars that cover a
significant range of musical styles turn out to be ambiguous
\cite{rohrmeier2011towards}. In most cases, this ambiguity is
eliminated by restricting the grammar to a specific genre, such as the
12-bar blues \cite{steedman1984generative} or other forms of jazz
\cite{harasim2018framework}.

In this paper, we propose to follow a different route. We considered a
fairly general grammar such as that in \cite{rohrmeier2011towards},
analyzed the various types of ambiguity, and looked for ways to remove
them. We concentrated in particular on \emph{modulation}, the process
of changing the key in the middle of a piece. Modulation cannot be
disambiguated in a context-free grammar and, rather than resorting to
a much more complex context-sensitive grammar, we use a numerical
method to detect modulation and integrate it into the analysis. The
result is, we believe, a better understanding of a process that has a
great impact on the effectiveness of the musical medium to convey
emotions and to highlight the contents of other media.

\section{Musical Background}
Harmony is the study of chords and their construction and chord
progressions and the principles of connection that govern them. In
this paper we examine the structural relation between
chords. Depending on the time period and the musical style these
principles may vary. Here, we focus on the Western classical music, in
particular on the tonal music of the XVII, XVIII and early XIX
century.

We can divide the harmony in 24 \emph{tonalities} (or \emph{keys}, we
shall use the two terms interchangeably). Each tonality is an
arrangement of notes, scales and chords with different functions and
relations. They relate with each other in a hierarchy where the tonic
chord is the one with the greatest stability and the rest of chords
play around it. Depending on the scales there are 12 major tonalities
(C, B$\flat$,...) and 12 minor tonalities (Cm,
B$\flat$m,...). Functionally, the major and minor mode of a tonality
are very similar and some composers of the XVIII and XIX century even
consider the two modes as two different aspects of the same
tonality. So we are going to consider that a tonality is the union of
its major and its minor modes (C is the union of C major and C minor),
considering a total of 12 keys.

Tonal music is based on diatonic scales, composed of seven notes
each. These notes are called the scale degrees, they are represented
with roman numerals ($I$,$II$,...,$VII$) and have different harmonic
functions. The main functions are \emph{tonic} ($I$), \emph{dominant}
($V$, $VII$), \emph{sub-dominant} ($IV$, $II$) and \emph{modal}
($III$, $VI$). Following standard musical theory, we consider the
latter as part of the tonic function.  The chords of a tonality are
built from the notes of the corresponding diatonic scales. In this
paper we use two different notations: the Roman numerals notation
($IV$ of C, or $IV_C$) and the English one (\textit{F}). The Roman
numerals one is the notation corresponded with the scale degree where
the chord is formed from, determining its function. This notation is
used along with the tonality to determine the chord. On the other
hand, the English notation determines a particular chord without
requiring the specification of a key.

Chords are played sequentially and each progression of chords can lead
to a more or less stable chord, creating musical interest. The
movement of a chord from an unstable sound to a more stable chord is
called a \emph{resolution}. The most common and stable resolutions are
called \emph{regular} resolutions, whereas an \emph{irregular}
resolutions are less common and less stable than a regular one.

The tonal center of a musical fragment is the note or chord with more
stability and it usually corresponds to the tonic of the key. We
mostly use chords near this tonal center, that is, chords that belong
to the corresponding key. But we can briefly leave the tonal center
(without leaving the tonality) in order to make music more
interesting, in order words, we can use chords that do not correspond
to the current key. An example of these chords are the secondary
dominants. As their name suggests, the secondary dominants are
dominant chords ($V$) of a scale degree that is not the tonic,
considering this scale degree as a new temporal tonic. For example, in
the key of C the secondary dominant of the chord $II$ (\textit{Dm}),
is the dominant of the key of Dm (\textit{A}), and we represent it as
$V^{II}$. The regular resolution of a secondary dominant is the chord
used to form it: $V^{II}$ - $II$.

Lastly, \emph{modulation} is the process of changing the tonality
within a musical fragment, that is to say, changing the tonal center
for a significant amount of time. There are different types of
modulation depending on the duration of the new tonal center and its
stability.

In essence, a tonality is formed by a determined set of chords. If we
only use those chords within a musical fragment, music will get
monotonous and lack of interest. For this reason, we usually change
from one tonality to another (modulation) and within a tonality we
occasionally use chords from other nearby tonalities (secondary
dominants).

\section{Grammar Analysis}
At the higher hierarchical level, a musical \emph{piece} is composed
of \emph{regions} that play, each one, one of the three major harmonic
roles: tonic (TR), dominant (DR), or sub-dominant
(SR). In order to avoid ambiguities in the parse tree, we define
separate terminals for the initial regions and their continuation (TR,
CTR, DR, CDR, SR, CSR). With these definitions, the top-level of our
grammar is:
\begin{equation}
  \begin{array}{rcl}
    \mbox{piece} & \rightarrow & TR \\
    TR & \rightarrow & CTR\ \ |\ \ CTR\ \ TR\ \ |\ \ CTR\ \ DR \\
    CTR & \rightarrow & DR\ \ t\ \ |\ \ t \\
    DR & \rightarrow & CDR\ \ |\ \ CDR\ \ DR \\
    CDR & \rightarrow & SR\ \ d\ \ |\ \ d \\
    SR & \rightarrow & CSR\ \ |\ \ CSR\ \ SR \\
    CSR & \rightarrow & s
  \end{array}
\end{equation}
These regions resolve in sub-trees rooted at the non-terminals $t$,
$d$, $s$, which represent the localized tonic, dominant, and
sub-dominant functions, respectively. The harmonic functions expand as
\begin{equation}
  \begin{array}{rcl}
    t & \rightarrow & dI\ \ |\ \ dI\ \ dIV\ \ dI\ \ |\ \ dVI\ \ |\ \ dIII \\
    s & \rightarrow & dIV\ \ |\ \ dII\ \ |\ \ \flat II \\
    d & \rightarrow & dV\ \ |\ \ dVII
  \end{array}
\end{equation}

These non-terminals are local harmonic functors that expand either to
the corresponding chord or to a secondary dominant with a regular 
(that resolves to that chord) or an irregular resolution. For 
example, $dIII$ resolves either to $III$ or to a secondary dominant
with its resolution:
\begin{equation}
 dIII \rightarrow III\ \ |\ \ V^{III}\ \ III\ \ |\ \ V^{V}\ \ III\ \ |\ \ VII^{III}\ \ III\ \ |\ \ VII^{V}\ \ III
\end{equation}
and similarly for the others. For example, if we are in the key of
$C$, this production would be equivalent to
\begin{equation}
    dIII \rightarrow Em\ \ |\ \ B\ \ Em\ \ |\ \ D\ \ Em\ \ |\ \ D\sharp \textsuperscript{o} \ \ Em\ \ |\ \ F\sharp \textsuperscript{o} \ \ Em
\end{equation}
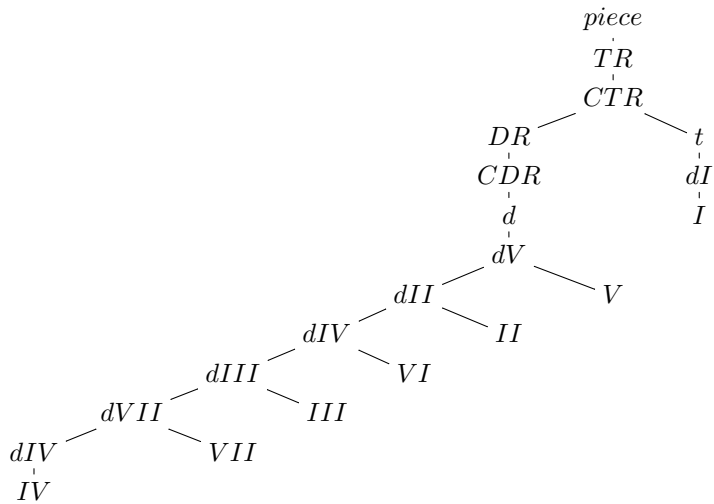
\begin{figure}[btp]
  \begin{center}
    $\displaystyle 
    \xymatrix@R=0.2em@C=1em{
      & & & & & & piece \ar@{-}[d]\\
      & & & & & & TR \ar@{-}[d]\\
      & & & & & & CTR \ar@{-}[dl] \ar@{-}[dr]\\
      & & & & & DR \ar@{-}[d] & & t \ar@{-}[d]\\
      & & & & & CDR \ar@{-}[d] & & dI \ar@{-}[d]\\
      & & & & & d \ar@{-}[d] & & I\\
      & & & & & dV \ar@{-}[dl] \ar@{-}[dr]\\
      & & & & dII \ar@{-}[dl] \ar@{-}[dr] & & V & & & & & \\
      & & & dIV \ar@{-}[dl] \ar@{-}[dr] & & II & & & &  \\
      & & dIII \ar@{-}[dl] \ar@{-}[dr] & & VI & & & & & & &  \\
      & dVII \ar@{-}[dl] \ar@{-}[dr] & & III & & & & & & \\
      dIV \ar@{-}[d]& & VII & & & & & & & & \\
      IV & 
    }$
  \end{center}
  \caption{\small Parse tree that results from the analysis of the
    sequence $IV\ VII\ III\ VI\ II\ V\ I$ using the descending
    fifths sequences productions.}
  \label{descending-fifths}
\end{figure}
The objective of this and the analogous expansions of $dI,$$\ldots,$ $dVII$
is to allow the introduction of secondary dominants at a limited
depth. Resolutions like $V^{III}\ \ III$ (the dominant of the 3rd
followed by the 3rd, or $B\ \ Em$ in the key of C) allows the
introduction of chords far from the tonal center of the piece, adding
interest to otherwise flat harmonies. 
In our grammar we have deliberately chosen to limit the depth of nesting of secondary
dominants. In \cite{rohrmeier2011towards}, for example, secondary dominants are
defined by a recursive rule, allowing constructions such as $V^V\ V$,
$V^{V^V}\ V^V\ V$, $V^{V^{V^V}}\ V^{V^V}\ V^V\ V$, and so on.
On the one hand, this results in an ambiguous grammar; on the other hand,
this allowed the specification of sequences like
$V^{V^{V^V}}\ V^{V^V}\ V^V\ V$ that, from the musical point of view,
make no sense as they go too far from the tonal center without
settling on a new one. Note that the secondary dominant rules are very
low in the hierarchy of productions, reflecting the fact that they are
local phenomena that do not span more than two or three chords.

Ambiguity also emerges when trying to introduce other common musical
constructs, such as the \emph{descending fifths}: a progression where
each chord is a fifth above the next one, all chords being in the same
key. For example, the sequence $Em\ Am\ Dm\ G\ C$ is a descending
fifths sequence in the key of C ($III\ VI\ II\ V\ I$).  These
progressions can in principle be analyzed using the productions of
\cite{rohrmeier2011towards}:
\begin{equation}
  \begin{array}{rcl}
    dI & \rightarrow & dV\ \ |\ \ I\\
    dII & \rightarrow & dVI\ \ |\ \ II\\
    dIII & \rightarrow & dVII\ \ |\ \ III\\
    dIV & \rightarrow & dI\ \ |\ \ IV\\
    dV & \rightarrow & dII\ \ |\ \ V\\
    dVI & \rightarrow & dIII\ \ |\ \ VI\\
    dVII & \rightarrow & dIV\ \ |\ \ VII
  \end{array}
  \label{desc-fifths-rules}
\end{equation}
obtaining parse trees as the one shown in Figure \ref{descending-fifths}.

\begin{figure}[tbp]
  \begin{center}
    $\displaystyle 
    \xymatrix@R=0.2em@C=0.5em{
      & & & & piece \ar@{-}[d] & &\\
      & & & & TR \ar@{-}[dll] \ar@{-}[drr] & &\\
      & & CTR \ar@{-}[dl] \ar@{-}[drr] & & & & TR \ar@{-}[dl] \ar@{-}[drr] &\\
      & DR \ar@{-}[d] & & & t \ar@{-}[d] & CTR \ar@{-}[d] & & & TR \ar@{-}[d]\\
      & CDR \ar@{-}[dl] \ar@{-}[dr] & & & tcp \ar@{-}[d] & t \ar@{-}[d] & & & CTR \ar@{-}[dl] \ar@{-}[dr]\\
      SR \ar@{-}[d] & & d \ar@{-}[d] & & dIII \ar@{-}[d] & tp \ar@{-}[d] & & DR \ar@{-}[d] & & t \ar@{-}[d]\\
      CSR \ar@{-}[d] & & dp \ar@{-}[d]& & III & dVI \ar@{-}[d] & & CDR \ar@{-}[dl] \ar@{-}[dr] & & dI \ar@{-}[d]\\
      s \ar@{-}[d] & & dVII \ar@{-}[d] & & & VI & SR \ar@{-}[d] & & d \ar@{-}[d] & I\\
      dIV \ar@{-}[d] & & VII & & & & CSR \ar@{-}[d] & & dV \ar@{-}[d]\\
      IV & & & & & & s \ar@{-}[d] & & V\\
      & & & & & & sp \ar@{-}[d]\\
      & & & & & & dII \ar@{-}[d]\\
      & & & & & & II
    }$
  \end{center}
  \caption{\small Parse tree that results from the analysis of the
    sequence $IV\ VII\ III\ VI\ II\ V\ I$ using the rest of the
    productions of the grammar, proving that the descending
    fifths sequence productions make the grammar ambiguous.}
  \label{descending-fifths-2}
\end{figure}
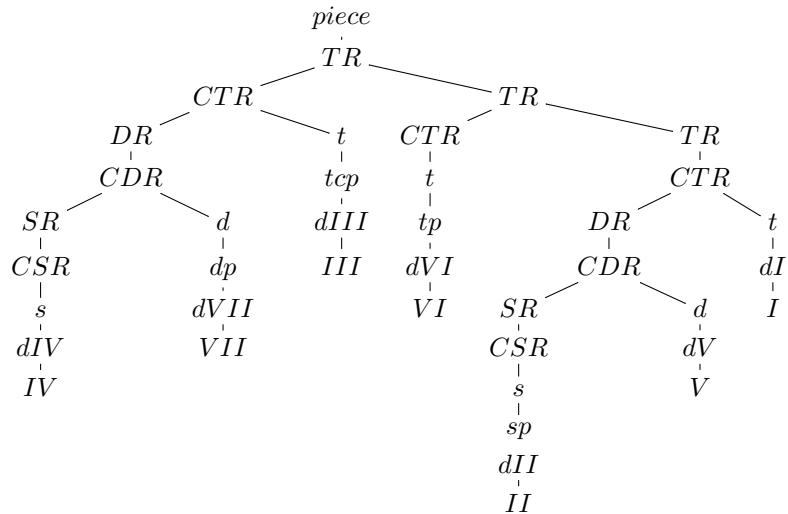

However, these sequences could be also analyzed using the rest of
productions of the grammar, as shown in Figure
\ref{descending-fifths-2}, which means that the rules of
\ref{desc-fifths-rules} make the grammar ambiguous. Furthermore, the
definition of this kind of sequences is, like many concepts in music,
a fuzzy one: it is not clear what is the minimum number of chords
needed to determine the existence of a descending fifths sequence, nor
whether it makes sense to talk of such minimum, independent of the
harmonic context in which the sequence is placed. For these reasons,
in this work we disregard descending fifths sequences, dropping the
corresponding productions from the grammar (see the conclusion for
more on this subject).

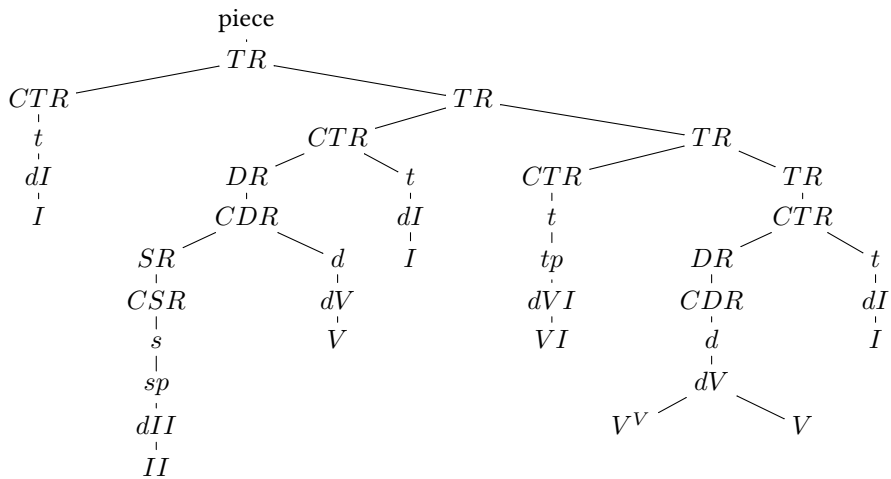
\begin{figure}[htbp]
  \begin{center}
    $\displaystyle 
    \xymatrix@R=0.2em@C=0.5em{
      & & & \mbox{piece} \ar@{-}[d] \\
      & & & TR \ar@{-}[dlll] \ar@{-}[drrr] \\
      CTR \ar@{-}[d] & & & & & & TR \ar@{-}[dll] \ar@{-}[drrr] \\
      t \ar@{-}[d] & & & & CTR \ar@{-}[dl] \ar@_{-}[dr] & & & & & TR \ar@{-}[dr] \ar@{-}[dll]\\
      dI \ar@{-}[d] & & & DR \ar@{-}[d] & & t \ar@{-}[d] & & CTR \ar@{-}[d] & & & TR \ar@{-}[d] \\
      I & & & CDR\ar@{-}[dr] \ar@{-}[dl]  & & dI \ar@{-}[d] & & t \ar@{-}[d] & & & CTR \ar@{-}[dr] \ar@{-}[dl] \\
       & & SR \ar@{-}[d] & & d \ar@{-}[d] & I & & tp \ar@{-}[d] & & DR \ar@{-}[d] & & t \ar@{-}[d] \\
      & & CSR \ar@{-}[d] & & dV \ar@{-}[d] & & & dVI \ar@{-}[d] & & CDR \ar@{-}[d] & & dI \ar@{-}[d] \\
       & & s \ar@{-}[d] & & V & & & VI & & d \ar@{-}[d] & & I \\
      & & sp \ar@{-}[d] & & & & & & & dV \ar@{-}[dl] \ar@{-}[dr]\\
      & & dII \ar@{-}[d] & & & & & & V^{V} & & V \\
      & & II
    }$
  \end{center}
  \caption{\small An example of harmonic analysis from the fragment
    $I\ II\ V\ I\ VI\ V^{V}\ V\ I$ from J. S. Bach's \emph{Prelude 1
      in C major} (BWV 846). The example shows a typical structure of
    tonic, dominant, and sub-dominant region, and a secondary dominant
    ($V^{V}$).}
  \label{bach1}
\end{figure}

The complete grammar is shown in
Figure~\ref{fullgrammar}. Figure~\ref{bach1} shows an example of an
analysis: we consider the sequence $I\ II\ V\ I\ VI\ V^{V}\ V\ I$ from
J. S. Bach's \emph{Prelude 1 in C major} (BWV 846).
\begin{figure*}[btp]
  \begin{center}
  {\small
    $\displaystyle
    \begin{array}{rclcrcl}
      \mbox{piece} & \rightarrow & TR & \rule{0em}{0pt}                 & dI & \rightarrow & V^{III}\ I\ |\ V^V\ I \\
      TR & \rightarrow & CTR\ |\ CTR\ TR\ |\ CTR\ DR    &    & dII & \rightarrow &\ V^{II}\ II\ |\ VII^{II}\ II\ |\ V^{IV}\ V^{II}\ II\\
      CTR & \rightarrow & DR\ t\ |\ t                         &   & dIII & \rightarrow & V^{III}\ III\ |\ V^V\ III\ |\ VII^{III}\ III\ |\ VII^V\ III\\
      DR & \rightarrow & CDR\ |\ CDR\ Dr & & dIV & \rightarrow & V^{IV}\ IV\ |\ V^{VI}\ IV\ |\ VII^{IV}\ IV\ |\ VII^{VI}\ IV\ |\ V^V\ V^{IV}\ IV\\
      CDR & \rightarrow & SR\ d\ |\ d                             & & dV & \rightarrow & V^V\ V\ |\ V^{II}\ V\ |\ VII^V\ V\ |\ VII^{II}\ V \\
      SR & \rightarrow & CSR\ |\ CSR\ SR                         & & dVI & \rightarrow & V^{VI}\ VI\ |\ VII^{VI}\ VI \\
      CSR & \rightarrow & s                                            & & dVII & \rightarrow & V^{VII}\ VII\ |\ VII^{VII}\ VII\\
      t & \rightarrow & tp\ |\ tcp\ dI\ |\ dI\ dV\ dI    & & dI & \rightarrow & I \\
      s & \rightarrow & sp \ |\ dIV                                & & dII & \rightarrow & II \\
      d & \rightarrow & dp\ |\ dV                                  & & dIII & \rightarrow & III \\
      tp & \rightarrow &\ dVI                                        & & dIV & \rightarrow & IV \\
      dp & \rightarrow &\ dVII                                       & & dV & \rightarrow & V \\
      sp & \rightarrow &\ dII\ |\ \flat{II}                      & & dVI & \rightarrow & VI \\
      tcp & \rightarrow & dIII                                         & & dVII & \rightarrow & VII 
    \end{array}
    $
  }
  \end{center}
  \caption{\small The grammar that we use in our analysis. Here, $I, II, III,
    IV, V, VI, VII$ are the terminals, while \dqt{piece} is the start
    symbol.}
  \label{fullgrammar}
\end{figure*}

\section{Modulation}
One of the major hurdles in grammar-based approaches is the detection
of modulation, that is, of the change of key in the middle of a
piece. Modulation is typically articulated around a \emph{pivot
  chord}, a chord that belongs to both keys and that marks the
transition point from one to the other:
\begin{center}
  \setlength{\unitlength}{1em}
  \begin{picture}(23,7)(0,0)
    \put(4,3){\makebox(0,0){C}}
    \put(6,3){\makebox(0,0){F}}

    \put(8,3){\makebox(0,0){G}}
    \put(10,3){\makebox(0,0){C}}
    \put(12,3){\makebox(0,0){Am}}
    \put(14,3){\makebox(0,0){D}}
    \put(16,3){\makebox(0,0){G}}
    \put(18,3){\makebox(0,0){Am}}
    \put(20,3){\makebox(0,0){D}}
    \put(22,3){\makebox(0,0){G}}    
    \put(3,2){\line(0,1){0.5}}
    \put(3,2){\line(1,0){10.2}}
    \put(13.2,2){\line(0,1){0.5}}
    \put(8,1.8){\makebox(0,0)[t]{\textbf{C}}}
    \put(0,1){\makebox(0,0)[l]{keys}}
    \put(11,1){\line(0,1){0.5}}
    \put(11,1){\line(1,0){12}}
    \put(23,1){\line(0,1){0.5}}
    \put(17,0.8){\makebox(0,0)[t]{\textbf{G}}}
    \put(12,5){\vector(0,-1){1.5}}
    \put(12,5.2){\makebox(0,0)[b]{pivot}}
  \end{picture}
\end{center}

\begin{figure}[hbtp]
  \centerline{\HInsert{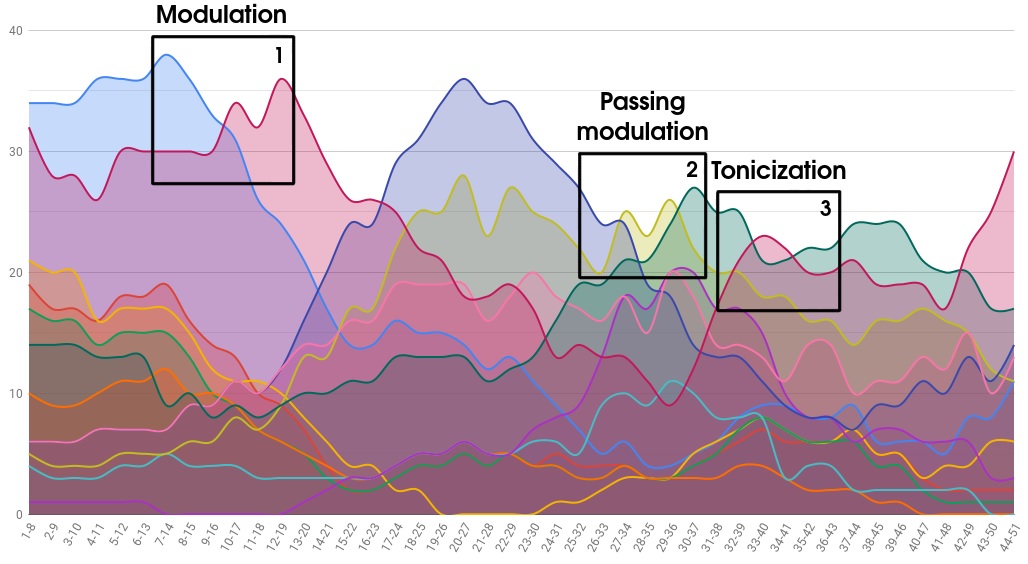}{30em}}
  \caption{\small Analysis of the modulation of Chopin's \emph{4
      Ballade}. The boxes indicate different types of modulation (see
    text). For the color codes of the curves and their correspondence
    with tonality, see Figure~\ref{ludwig}.}
  \label{modulations}
\end{figure}

This process is inherently ambiguous for two reasons. First,
each chord has a harmonic interpretation in any key, so that sequences
of chords can be interpreted in exponentially many ways. In
Figure~\ref{interpret}, for example, we interpret the same sequence of
chords with a modulation (the most natural way, from the point of view
of music theory), without modulation (in a single key), and with an
arbitrary sequence of modulations (the subscript is the key in which
we interpret the chord, resulting in the harmonic function indicated).
\begin{figure}[htbp]
\begin{center}
  \setlength{\unitlength}{1em}
  \begin{picture}(24,3)(0,0)
    \put(-1,3){%
      \put(4,0){\makebox(0,0){\textbf{C}}}
      \put(6,0){\makebox(0,0){\textbf{F}}}
      \put(8,0){\makebox(0,0){\textbf{G}}}
      \put(10,0){\makebox(0,0){\textbf{C}}}
      \put(13.5,0){\makebox(0,0){\textbf{Am}}}
      \put(17,0){\makebox(0,0){\textbf{D}}}
      \put(19,0){\makebox(0,0){\textbf{G}}}
      \put(21,0){\makebox(0,0){\textbf{Am}}}
      \put(23,0){\makebox(0,0){\textbf{D}}}
      \put(25,0){\makebox(0,0){\textbf{G}}}
    }
    \put(-1,2){%
      \put(04,0){\makebox(0,0){$I_C$}}
      \put(06,0){\makebox(0,0){$IV_C$}}
      \put(08,0){\makebox(0,0){$V_C$}}
      \put(10,0){\makebox(0,0){$I_C$}}
      \put(13.5,0){\makebox(0,0){$VI_C/II_G$}}
      \put(17,0){\makebox(0,0){$V_G$}}
      \put(19,0){\makebox(0,0){$I_G$}}
      \put(21,0){\makebox(0,0){$II_G$}}
      \put(23,0){\makebox(0,0){$V_G$}}
      \put(25,0){\makebox(0,0){$I_G$}}    
    }
    \put(-1,1){%
      \put(04,0){\makebox(0,0){$I_C$}}
      \put(06,0){\makebox(0,0){$IV_C$}}
      \put(08,0){\makebox(0,0){$V_C$}}
      \put(10,0){\makebox(0,0){$I_C$}}
      \put(13.5,0){\makebox(0,0){$VI_C$}}
      \put(17,0){\makebox(0,0){$V_C^{V}$}}
      \put(19,0){\makebox(0,0){$V_C$}}
      \put(21,0){\makebox(0,0){$VI_C$}}
      \put(23,0){\makebox(0,0){$V_C^{V}$}}
      \put(25,0){\makebox(0,0){$V_C$}}    
    }
    \put(-1,0){%
      \put(04,0){\makebox(0,0){$V_F$}}
      \put(06,0){\makebox(0,0){$I_F$}}
      \put(08,0){\makebox(0,0){$III_E$}}
      \put(10,0){\makebox(0,0){$VI_E$}}
      \put(13.5,0){\makebox(0,0){$IV_E$}}
      \put(17,0){\makebox(0,0){$I_D$}}
      \put(19,0){\makebox(0,0){$V_D$}}
      \put(21,0){\makebox(0,0){$I_{Am}$}}
      \put(23,0){\makebox(0,0){$V_G$}}
      \put(25,0){\makebox(0,0){$I_G$}}    
    }
    \put(0,2){\makebox(0,0)[l]{(A)}}
    \put(0,1){\makebox(0,0)[l]{(B)}}
    \put(0,0){\makebox(0,0)[l]{(C)}}
  \end{picture}
\end{center}
  \caption{\small A sequence of chords interpreted in three different ways:
    with a modulation (in (A), the most natural way, from the point of
    view of music theory), without modulation (in (B), in a single
    key), and with an arbitrary sequence of modulations (in (C)). The
    subscripts represent the key in which we are interpreting the
    chord to give it the indicated harmonic function.}
  \label{interpret}
\end{figure}
Second, the presence of a chord far away from the tonal center is not
enough to define a modulation: it is necessary to have a reasonably
long sequence of chords reasonably close to a different tonal
center. The double use of \dqt{reasonably} shows that the definition
is fuzzy, and not easy to analyze with a formal grammar, at least, not
with a context-free one. We have therefore decided to decouple the
analysis of the modulation from the grammar. Modulation is
detected by gathering evidence on the local key by giving each chord a
score that depends on the centrality of that chord. For example,
Table~\ref{Ckey} shows the centrality of various chords for the
key of C.
\begin{table*}[thbp]
  \begin{center}
    \begin{tabular}{|c|l|c|c|c|c|c|c|c|c|}
      \cline{3-10} 
      \multicolumn{2}{c|}{} & $I$ & $II$ & $III$ & $IV$ & $V$ & $VI$ & $VII$ & $\flat{II}$ \\
      \hline
      \multirow{2}{*}{\textbf{C}} & major & C (5) & Dm (3) & Em (2) & F (3) & G/G7 (5) & Am (3) & B\textsuperscript{o} (3) &  \multirow{6}{*}{D$\flat$ (1)} \\
      \cline{2-9}
      & secondary dominant & - & A/A7 (1) & B/B7 (1) & C7 (1) & D/D7 (1) & E/E7 (1) & - & \\
      \cline{1-9}
      \multirow{4}{*}{\textbf{Cm}}& ascending melodic & - & Dm (3) & - & F (3) & - & - & - &  \\
      \cline{2-9}
      & harmonic minor & Cm (5) & D\textsuperscript{o} (2) & - & Fm (3) & G/G7 (5) & A$\flat$ (2) & B\textsuperscript{o} (3) & \\
      \cline{2-9}
      & descending melodic & - & - & E$\flat$ (2) & - & Gm (2) & - & B$\flat$ (2) & \\
      \cline{2-9}
      & secondary dominant & - & - & B$\flat$7 (1) & C7 (1) & D/D7 (1) & E$\flat$7 (1) & F7 (1) & \\
      \hline
    \end{tabular}
  \end{center}
  \caption{\small Table with the scores that determine the chord centrality
    for the key of C. Chords not represented in the table have a
    score of 0.}
  \label{Ckey}
\end{table*}
\begin{table}[tbhtp]
  \begin{center}
    \begin{tabular}{|c|c|c|c|c|c|c|c|c||c|}
      \cline{2-10} 
      \multicolumn{1}{c|}{} & C & F & C & Dm & G7 & C & G7 & C & Tot. \\
      \hline
      \textbf{C} & 5 & 3 & 5 & 3 & 5 & 5 & 5 & 5 & \textbf{36} \\
      \hline
      \textbf{G} & 3 & 2 & 3 & 2 & 1 & 3 & 1 & 3 & 18 \\
      \hline
      \textbf{D} & 2 & 2 & 2 & 5 & 1 & 2 & 1 & 2 & 17 \\
      \hline
      \textbf{A} & 2 & 2 & 2 & 3 & 1 & 2 & 1 & 2 & 15 \\
      \hline
      \textbf{E} & 2 & 1 & 2 & 0 & 1 & 2 & 1 & 2 & 11 \\
      \hline
      \textbf{B} & 1 & 0 & 1 & 0 & 0 & 1 & 0 & 1 & 4 \\
      \hline
      \textbf{F$\sharp$} & 0 & 1 & 0 & 0 & 0 & 0 & 0 & 0 & 1 \\
      \hline
      \textbf{C$\sharp$} & 1 & 1 & 1 & 0 & 0 & 1 & 0 & 1 & 5 \\
      \hline
      \textbf{A$\flat$}  & 1 & 1 & 1 & 0 & 1 & 1 & 1 & 1 & 7 \\
      \hline
      \textbf{E$\flat$}  & 1 & 1 & 1 & 0 & 1 & 1 & 1 & 1 & 7 \\
      \hline
      \textbf{B$\flat$}  & 1 & 5 & 1 & 2 & 1 & 1 & 1 & 1 & 13 \\
      \hline
      \textbf{F} & 5 & 5 & 5 & 3 & 1 & 5 & 1 & 5 & 30 \\
      \hline
    \end{tabular}
  \end{center}
  \caption{\small Determination of the key of the sequence
    $C\ F\ G\ Dm\ G7\ C\ G7\ C$ (top row). Each chord is scored in all
    keys and the scores are accumulated (last column). The highest
    score indicates that the fragment is in the key of C.}
  \label{chordscore}
\end{table}
There are 12 such tables, one for each key. In order to determine the
key of a sequence of chords, we consider the score of each chord in
each of the 12 keys and accumulate the scores: the key with the
highest score will be considered as the key of the sequence of
chords. Consider, for example, the sequence
$C\ F\ G\ Dm\ G7\ C\ G7\ C$. Table~\ref{chordscore} shows the process
of detecting the key: each column represents the score of that chord
in each of the keys (e.g., $C$ has a weight of $5$ in the key of C,
$3$ in the key of G, $2$ in the key of D, etc.). At the end of each
row we indicate the accumulated score for the corresponding key. The
highest score indicates that the fragment is in the key of C.  Based
in this general idea, modulation is detected by sliding a window on
the musical piece and detecting, in each window, the dominant
key. When the dominant key changes, we detect a modulation, and the
central chord of the window in which the key changes is taken as the
pivot. Table~\ref{modulate} shows the analysis of the sequence
$C\ F\ C\ Dm\ G7\ C\ G7\ C\ C\ C\ Fm\ E^{o}\ Fm\ E^{o}\ Fm$; When the
window covers chords 6 to 13, a key change is
\begin{table}[thbp]
  \begin{center}
    \begin{tabular}{|c|c|c|c|c|c|c|c|c|}
      \cline{2-9} 
      \multicolumn{1}{c|}{} & 1-8 & 2-9 & 3-10 & 4-11 & 5-12 & 6-13 & 7-14 & 8-15 \\
      \cline{1-9}
      \textbf{C} & \textbf{36} & \textbf{36} & \textbf{38} & \textbf{36} & \textbf{33} & 31 & 26 & 24 \\
      \cline{1-9} 
      \textbf{G} & 18 & 18 & 19 & 16 & 14 & 13 & 10 & 9 \\
      \cline{1-9}
      \textbf{D} & 17 & 17 & 17 & 15 & 12 & 11 & 11 & 10 \\
      \cline{1-9} 
      \textbf{A} & 15 & 15 & 15 & 13 & 10 & 9 & 7 & 6 \\
      \cline{1-9}
      \textbf{E} & 11 & 11 & 12 & 10 & 10 & 9 & 7 & 6 \\
      \cline{1-9}
      \textbf{B} & 4 & 4 & 5 & 4 & 4 & 4 & 3 & 3 \\
      \cline{1-9}
      \textbf{F$\sharp$} & 1 & 1 & 0 & 0 & 0 & 0 & 0 & 0 \\
      \cline{1-9}
      \textbf{C$\sharp$} & 5 & 5 & 5 & 6 & 6 & 8 & 7 & 9 \\
      \cline{1-9}
      \textbf{A$\flat$} & 7 & 7 & 7 & 9 & 9 & 11 & 10 & 12  \\
      \cline{1-9}
      \textbf{E$\flat$} & 7 & 7 & 7 & 9 & 9 & 11 & 10 & 12 \\
      \cline{1-9}
      \textbf{B$\flat$} & 13 & 13 & 9 & 10 & 8 & 9 & 8 & 9  \\
      \cline{1-9}
      \textbf{F} & 30 & 30 & 30 & 30 & 30 & \textbf{34} & \textbf{32} & \textbf{36} \\
      \cline{1-9}
      \multicolumn{1}{c|}{} & \textbf{C} & \textbf{C} & \textbf{C} & \textbf{C} & \textbf{C} &
                              \textbf{F} & \textbf{F} & \textbf{F} \\
      \cline{2-9}
    \end{tabular}
  \end{center}
  \caption{\small Determination of the modulation of a sequence using a
    sliding window of 8 chords. Each column represents the accumulated
    scores of the 12 keys for the position of the window shown in the
    first row. The last row shows the dominant key for each position
    of the window. A modulation is detected in window 6-13, thus
    identifying the 9th chord (position $\lfloor(13-6+1)/2\rfloor$ of the window) as the pivot.}
  \label{modulate}
\end{table}
detected. Correspondingly, we place the pivot chord in position 9
(position $\lfloor(13-6+1)/2\rfloor$ of the window), the central $C$ of the series of three:
\begin{center}
  \setlength{\unitlength}{0.8em}
\begin{picture}(32,6)(0,0)
  \put(2,4){\line(0,1){0.5}}
  \put(20,4){\line(0,1){0.5}}
  \put(2,4){\line(1,0){18}}
  \put(18,2){\line(0,1){0.5}}
  \put(32,2){\line(0,1){0.5}}
  \put(18,2){\line(1,0){14}}
  \put(3,5){\makebox(0,0){C}}
  \put(5,5){\makebox(0,0){F}}
  \put(7,5){\makebox(0,0){C}}
  \put(9,5){\makebox(0,0){Dm}}
  \put(11,5){\makebox(0,0){G7}}
  \put(13,5){\makebox(0,0){C}}
  \put(15,5){\makebox(0,0){G7}}
  \put(17,5){\makebox(0,0){C}}
  \put(19,5){\makebox(0,0){\textbf{C}}}
  \put(21,5){\makebox(0,0){C}}
  \put(23,5){\makebox(0,0){Fm}}
  \put(25,5){\makebox(0,0){E\textsuperscript{o}}}
  \put(27,5){\makebox(0,0){Fm}}
  \put(29,5){\makebox(0,0){E\textsuperscript{o}}}
  \put(31,5){\makebox(0,0){Fm}}
  \put(3,3){\makebox(0,0){I}}
  \put(5,3){\makebox(0,0){IV}}
  \put(7,3){\makebox(0,0){I}}
  \put(9,3){\makebox(0,0){II}}
  \put(11,3){\makebox(0,0){V7}}
  \put(13,3){\makebox(0,0){I}}
  \put(15,3){\makebox(0,0){V7}}
  \put(17,3){\makebox(0,0){I}}
  \put(19,3){\makebox(0,0){I}}
  \put(19,1){\makebox(0,0){V}}
  \put(21,1){\makebox(0,0){V}}
  \put(23,1){\makebox(0,0){I}}
  \put(25,1){\makebox(0,0){VII}}
  \put(27,1){\makebox(0,0){I}}
  \put(29,1){\makebox(0,0){VII}}
  \put(31,1){\makebox(0,0){I}}
  \put(0,1){\makebox(0,0){\textbf{F}}}
  \put(0,3){\makebox(0,0){\textbf{C}}}
\end{picture}

\end{center}
Different types of modulations are detected as different patterns of
score changes in the sliding window. The most common examples are
shown in Figure~\ref{modulations}.
In Box 1 we have a regular modulation: the key of the piece changes
from C to F; both the key of C (prior to the modulation) and that of F
(afterwards) are maintained for a significant amount of time and are
well established to the ear. In Box 2 we have an example of
\emph{passing} (or \emph{false}) \emph{modulation}: we have a sequence of three keys, A$\flat$,
D$\flat$, and B$\flat$. A$\flat$ and B$\flat$ are dominant for some
time, and are thus well established, but D$\flat$ is dominant for a
short time, never becoming well established and working only as a
nexus between A$\flat$ and B$\flat$. Finally, in Box 3, we have an
example of \emph{tonicization}: here too we
have a key (F) that stays dominant for a short period but, in this
case, before and after the modulation we are in the same key:
B$\flat$. This solution is used by composers when they want to add
chromaticism to the music: here, F is the dominant of B$\flat$ and in
that fragment there are chords that are related to F, not with the
purpose of changing key, but to have a more \dqt{dominantly-colored}
harmony on B$\flat$.

\section{Results}
The most important parameter of our method is the window length
$W$. Large values of $W$ ($W\sim{10-12}$) give a more firm
establishment of the dominant key, as they allow the method to gather
more evidence. On the other hand, large values of $W$ act as a
low-pass filter on the sequence of chords, which could lead to
overlook fast tonicizations or false modulations. Short windows sizes
($W\sim{2-4}$), on the other hand, could lead to over-detection of
tonicization simply due to the presence of a few chords far from the
tonal center.
\begin{figure*}[btp]
  \begin{center}
        \begin{tabular}{ccc}
          & \rule{1em}{0pt} & \\
          \HInsert{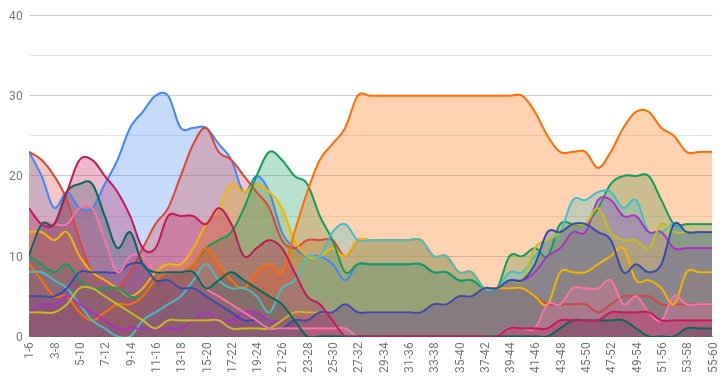}{20em} & &
          \HInsert{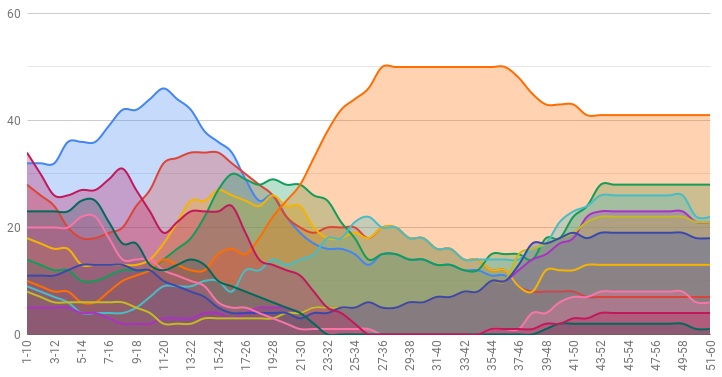}{20em}  \\
          \multicolumn{3}{c}{
            \HInsert{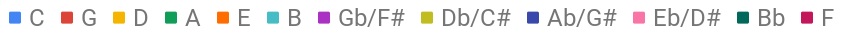}{40em}}
        \end{tabular}
  \end{center}
  \caption{\small Analysis of the modulation of Beethoven's \emph{Waldstein
      Sonata} using windows length $W=6$ (left) and $W=10$ (right).}
  \label{ludwig}
\end{figure*}
The problem is made more complicated by the fact that there is, in
these cases, no established ground truth: whether a series of chords
are a tonicization (or a false modulation) or simply a few chords far
\begin{figure}[thb]
  \begin{center}
    \setlength{\unitlength}{0.76em}
\begin{picture}(32,22)(0,-1)
  \put(3,4){\line(1,0){2}}
  \put(8,4){\line(1,0){3}}
  \put(14,4){\line(1,0){3}}
  \put(20,4){\line(1,0){3.5}}
  \put(25.5,4){\line(1,0){4}}
  \put(31.5,4){\line(1,0){2.5}}
  \put(3,8){\line(1,0){2}}
  \put(8,8){\line(1,0){3}}
  \put(14,8){\line(1,0){20}}
  \put(3,12){\line(1,0){2}}
  \put(8,12){\line(1,0){26}}
  \put(3,16){\line(1,0){2}}
  \put(8,16){\line(1,0){26}}
  \put(3,20){\line(1,0){31}}
  \thicklines
  \put(3,0){\line(1,0){31}}
  \put(5,0){\line(0,1){19.636}}
  \put(8,0){\line(0,1){19.636}}
  \put(5,19.636){\line(1,0){3}}
  \put(6.8,14.5){\makebox(0,0)[t]{\begin{rotate}{90}\textbf{49.09\%}\end{rotate}}}
  \put(11,0){\line(0,1){10.908}}
  \put(14,0){\line(0,1){10.908}}
  \put(11,10.908){\line(1,0){3}}
  \put(12.8,6){\makebox(0,0)[t]{\begin{rotate}{90}\textbf{27.27\%}\end{rotate}}}
  \put(17,0){\line(0,1){8}}
  \put(20,0){\line(0,1){8}}
  \put(17,8){\line(1,0){3}}
  \put(18.8,2.8){\makebox(0,0)[t]{\begin{rotate}{90}\textbf{20.00\%}\end{rotate}}}
  \put(23,0){\line(0,1){0.728}}
  \put(26,0){\line(0,1){0.728}}
  \put(23,0.728){\line(1,0){3}}
  \put(24.8,1.2){\makebox(0,0)[t]{\begin{rotate}{90}\textbf{1.82\%}\end{rotate}}}
  \put(29,0){\line(0,1){0.728}}
  \put(32,0){\line(0,1){0.728}}
  \put(29,0.728){\line(1,0){3}}
  \put(30.8,1.2){\makebox(0,0)[t]{\begin{rotate}{90}\textbf{1.82\%}\end{rotate}}}
  \put(0,-0.3){%
    \put(6.5,0){\makebox(0,0)[t]{0}}
    \put(12.5,0){\makebox(0,0)[t]{1}}
    \put(18.5,0){\makebox(0,0)[t]{2}}
    \put(25.5,0){\makebox(0,0)[t]{3}}
    \put(30.5,0){\makebox(0,0)[t]{4}}
    \put(17,-1){\makebox(0,0)[t]{distance from true pivot}}
  }
  \put(2.5,0){%
    \put(0,0){\makebox(0,0)[r]{0}}
    \put(0,4){\makebox(0,0)[r]{10\%}}
    \put(0,8){\makebox(0,0)[r]{20\%}}
    \put(0,12){\makebox(0,0)[r]{30\%}}
    \put(0,16){\makebox(0,0)[r]{40\%}}
    \put(0,20){\makebox(0,0)[r]{50\%}}
    \put(-3,6){\makebox(0,0){\begin{rotate}{90}\% of samples\end{rotate}}}
  }
\end{picture}
  \end{center}
  \caption{\small Precision in the placement of the pivot chords, from a
    sample of 55 modulations. In about 97\% of the cases the method
    places the pivot chord within two chords of the theoretical one
    determined by harmonic analysis.}
  \label{pivot}
\end{figure}
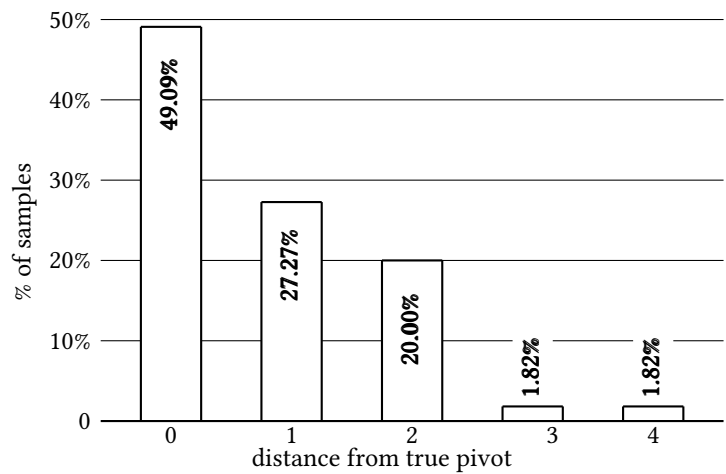
from the tonal center, is a matter of interpretation, and different
analysts may (and will) differ in their analysis.

An important factor in determining the value of $W$ is the style of
the music that we are analyzing. We have observed that a value of
$W=6$ is adequate for music from the XVII to the middle of the XIX
century. Modern music, up to the early XX century, is more
experimental with key variations, and authors do resort to rapid, short
changes of key to the point that sometimes the very concept of tonal
center is lost. In these cases, smaller values of $W$ must be
used. Large values of $W$ are useful also when we want to ignore the
rapid variations of key and detect the main harmonic themes of a piece.

The beginning of Beethoven's \emph{Waldstein Sonata} is an example
(Figure~\ref{ludwig}): the key values oscillate very rapidly, and do
not settle on any established key---the fragment is harmonically
ambiguous. With a longer value of $W$ (Figure~\ref{ludwig}b) we can
appreciate that the dominant key of the beginning of the Waldstein
Sonata is C.
An important measure to evaluate our method is the precision with
which we can identify the pivot chord. Figure~\ref{pivot}
shows the distance between the position of the pivot chord as
estimated by our method and that determined by musical analysis. The
data are based on a sample of 55 modulations from musical pieces of
the XVII, XVIII, and XIX centuries. In about 97\% of the cases, the
estimated pivot is within two chords of the actual one.

\begin{figure*}[bhtp]
  \begin{center}
    \input{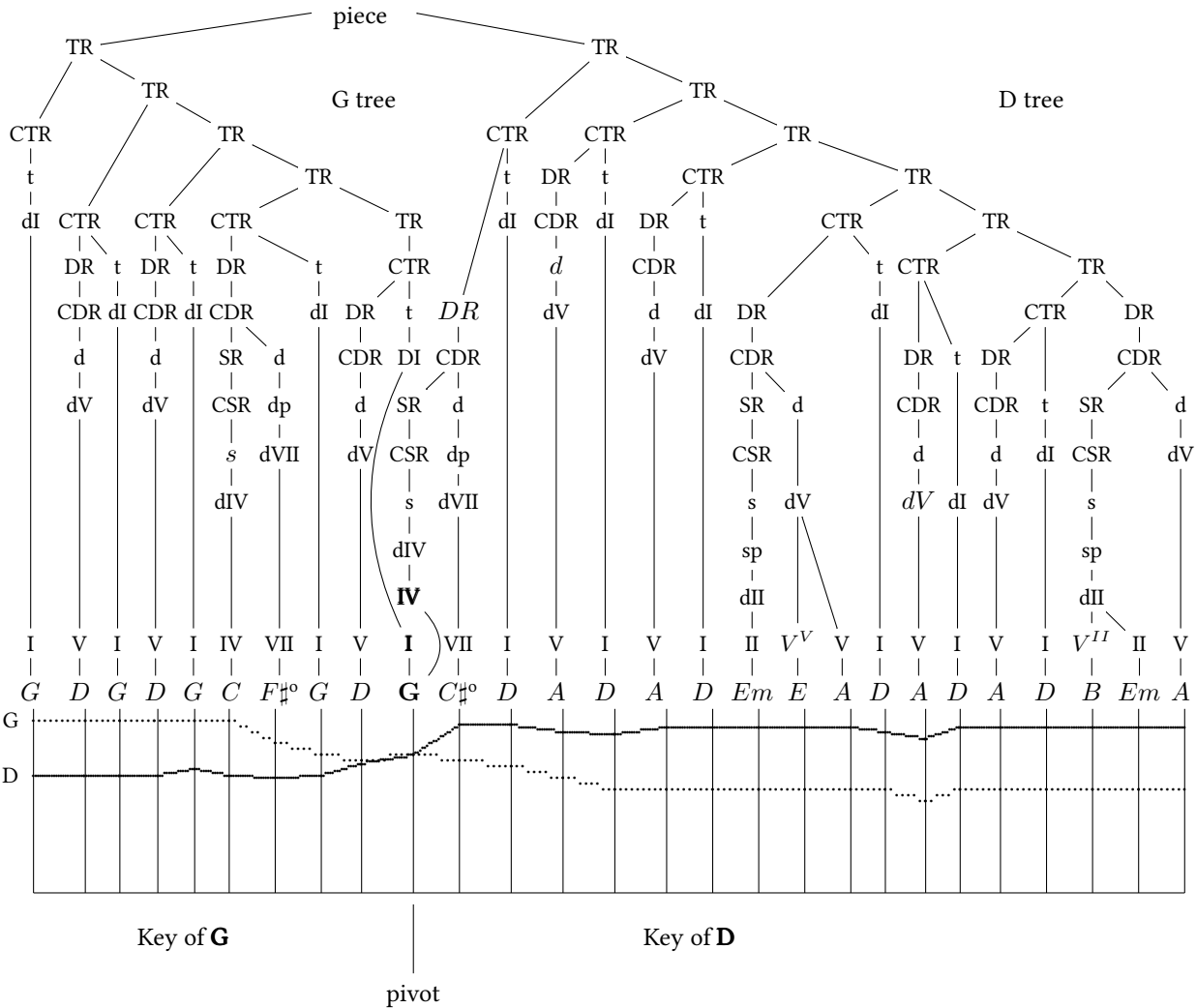}
  \end{center}
  \caption{\small Complete analysis of Mozart's \emph{Eine kleine
      Nachtmusik}. At the bottom of the figure we show the scores at
    the various chords of the music. For the sake of clarity, we only
    show the scores of the keys of \textbf{G} and \textbf{D}. The
    chord indicated as \emph{pivot} marks the change in key from
    \textbf{G} to \textbf{D}. The top part of the figure contains the
    trees that analyze the two parts. The harmonic functions (line
    just above the chords) have been assigned considering the chord
    and the key using a simple correspondence table. Note that the
    pivot is assigned two harmonic functions, as it is analyzed in
    both keys: it is $I$ (the tonic) in the key of \textbf{G} and $IV$
    (the sub-dominant) in the key of \textbf{D}. This causes that both
    trees have a branch ending in that chord.}
  \label{bigone}
\end{figure*}

Finally, in Figure~\ref{bigone} we show a complete analysis of a
musical fragment, from Mozart's \emph{Eine kleine Nachtmusik}.
At the bottom we show the weights of the main keys involved (we do not
show the score of all the 12 keys for the sake of clarity: the other
keys receive a score well below the two we show): the piece begins in
G and, after 10 chords, switches to D. The $G$ chord marked in bold is
the pivot, and belongs to both parts; consequently, it has two
different harmonic functions: it is a tonic (I) for the key of G and a
sub-dominant (IV) for the key of D. Each key is analyzed by a separate
tree, and the two tree have one leaf in common: the pivot.

\section{Conclusions}
The original objective of this investigation was to use formal
grammars to analyze the harmonic structure of a musical fragment,
studying their scope and limitations. Nevertheless, during the project
we discovered that most of the limitations of the grammar come from
the inherent ambiguous nature of music. Thus, it was not possible to
adapt the grammar to analyze these ambiguous musical
concepts. Particularly, we studied deeply the notion of modulation
and, after concluding that formal context-free grammars could not
represent it, we decided to use other methods to analyze modulations
in order to complement the formal grammars for the harmonic
analysis. Due to the different functions and importances of chords
within the different tonalities, we decided to use a
evidence-gathering numerical method to detect the different tonalities
of a musical fragment and therefore the modulations.

We ended with a hybrid method where we first detect the modulations
with the numerical method and then, based on the detected tonalities,
we analyze their harmonic structure using the formal grammars. The
results were really successful as we could analyze different harmonic
properties using the methods that better fit each one and combine them
obtaining really accurate analysis. This mixed approach could be
extended using the methods described on this paper or other methods
(as Markov chains, genetic algorithms, machine learning, etc.) to
analyze other ambiguous musical concepts. Hence, this is a very
promising research area that could allow us to have more powerful
analysis tools.

With regard to the numerical method for detecting mo\-dulations, it
allowed us to detect tonalities with a small error using the context:
how well the chords were combined with each other in each of the 12
tonalities. This is a very encouraging method that could be adapted to
analyze other musical properties that depend deeply on the context, as
the descending fifths and harmonic sequences. We would build up
evidence for the various competing hypothesis and adopt the solution
best supported by the evidence. Once the sequence has been divided
into unambiguous parts, grammars can be used to analyze them.

It is important to recall that the musical target of the investigation
was the tonal music of the XVII, XVIII and early XIX century,
developed mostly in Europe. The harmony of this music is organized
around tonalities and follows a particular set of rules that differ
from other musical styles. This makes the developed methods of this
project fail when they analyze musical fragments from styles that are
very different from the chosen one. The grammar rules are based on the
common chord progressions and relations of tonal music, so it cannot
properly analyze musical fragments that follow other relations, as
jazz music, and even that have different chords, as Arab
music. Moreover, the numerical method for detecting modulations is
based on the tonalities and hierarchy of chords in them, so applying
it to other musical styles, as modal music (or quasi-modal music, as
flamenco), does not make any sense, as modal music is based on modes
instead of tonalities. Thus, the method was thought and works great
with tonal music, with a certain melodic and harmonic structure.

Lastly, it is important to highlight the usefulness of understanding
the harmonic structure that underlies music and the importance that
harmonic analysis has in music. This way, we are able to know how
music, as media, communicates and therefore we could be able to use it
better along other mediums such as video and audio. It is not only a
way of better understanding the nature of music and feelings, it could
also be applied in the process of automatic generation of music and we
could combine it better with other multimedia systems.

\end{document}